\documentstyle[12pt]{article}
\def\dspace{\baselineskip = .30in}
\def\strut{\rule[-.5cm] {0cm}{1cm}}
\begin{document}

\title{MSSM FROM SUSY TRINIFICATION}

\author{{\bf G. Lazarides and C. Panagiotakopoulos}\\ Physics
Division\\School of Technology\\ University of Thessaloniki\\
Thessaloniki, Greece}

\date { }
\maketitle

\dspace
\centerline{\bf Abstract}
\vspace{.2in}

We construct a $SU(3)^3$ supersymmetric gauge theory with a common
gauge coupling g using only fields belonging to 27, $\overline {27}$
and singlet representations of $E_6$.
Spontaneous breaking of this gauge group at a scale
$M_X=1.3\times10^{16} $GeV gives naturally rise exactly to the Minimal
Supersymmetric Standard Model $(MSSM)$ and consequently to the
experimentally favored values of  $sin^2\theta_w$ and $\alpha_s$.The
gauge hierarchy problem is naturally solved by a missing-partner-type
mechanism which works to all orders in the superpotential. The baryon
asymmetry can be generated in spite of the (essential) stability of
the proton. The solar neutrino puzzle is solved by the  MSW mechanism.
The LSP is a natural "cold" dark matter candidate  and "hot" dark
matter might consist of $\tau$-neutrinos. This model could be thought
of as an effective $4d$ theory  emerging from a more fundamental theory
at a scale $M_c=M_P/\sqrt{8\pi}$ where $a_G\equiv{g^2\over{4\pi}}$
happens to be  equal to unity.
\newpage

The Minimal Supersymmetric Standard Model (MSSM) with a supersymmetry
breaking scale $M_s \sim 1 TeV$ is remarkably consistent with
unification of the three gauge coupling constants at a scale $\sim
10^{16} GeV$ assuming the  observed values for $sin^2 \theta_w $ and
$\alpha_s$ at the electroweak scale. This remarkable fact has been widely
advertized $^{(1)}$ as evidence for supersymmetry in connection with the
minimal supersymmetric $SU(5)$ model$^{(2)}$ which predicts exactly
the
spectrum of the $MSSM$ below a unification scale $M_X \sim 10^{16}
GeV$.

Despite this undoubted success, however, it appears that the minimal
$SU(5)$ scheme faces several difficulties. For instance, the predicted
asymptotic mass relations involving the first two families seem to be
in contradiction with the  observations. Furthermore, the close
relationship between the light fermion masses and proton decay through
$d=5$ operators  together with the lack of observation of  such events
could soon lead to exclusion of the minimal scheme. Additional reasons
for abandoning the minimal $SU(5)$ framework come from considerations
of neutrino masses, baryon asymmetry, etc. Finally, one should not
forget the gauge hierarchy problem.

In a recent paper$^{(3)}$ an effort has been made to overcome some
of the difficulties of the
minimal supersymmetric $SU(5)$ model by embedding the standard group
$SU(3)_c\times SU(2)_L \times U(1)_Y$ in $G\equiv SU(3)_c \times
SU(3)_L \times SU(3)_R$. This gauge group, which often arises in the simplest
compactification schemes of the heterotic $E_8 \times E_8$
superstring theory, has several advantages. Leptons
and quarks belong to separate representations which results in
avoiding wrong mass relations between light fermions. For the same
reason, one avoids the close relationship between the light quark
masses and the proton decay amplitude through $d=5$ interactions
encountered  in minimal unification schemes like $SU(5)$ or $SO(10)$.
The imposition of a simple $Z_2$ symmetry proved sufficient  to
guarantee the almost complete absence of baryon number violation in
the perturbative sector of the theory. The observed baryon asymmetry
was then generated by combining a leptogenesis mechanism with  the
non-perturbative baryon-and-lepton-number-violating effects of the
standard electroweak theory. In this first attempt specific models  were
constructed which were "close" to the MSSM below a unification scale
$M_X \sim 10^{16} GeV$, but the  gauge hierarchy problem was not
addressed at all.

The purpose of the present note is to construct a specific model based
on the gauge group $G\equiv SU(3)_c \times SU(3)_L \times SU(3)_R$
which below a unification scale $ M_X \sim 10^{16}$ GeV gives
naturally rise exactly to the MSSM. This guarantees the successful "
predictions" of the MSSM for $sin^2\theta_w$ and $\alpha_s$ providing, in
addition, a reason for the asymptotic equality of the three gauge
coupling constants. We assume that the theory emerges as a low energy
effective theory from a more involved construction at a scale $M_c$
close to the Planck mass  $M_P\simeq 1.2 \times 10^{19} GeV$. At the
scale $M_c$, the three $SU(3)$ gauge couplings are equal and, due to a
symmetric spectrum among them, they remain equal (in the one loop
approximation)
until the scale $M_X
\sim 10^{16} GeV$ where G breaks down to the standard gauge group.
Using appropriate discrete symmetries we forbid (essentially
completely) proton decay and we succeed in solving the gauge hierarchy
problem. Actually, we obtain only one light pair of electroweak higgs
doublets which does not acquire any  mass from renormalizable  as well
as non-renormalizable superpotential terms as long as an anomalous
$Z_2$  discrete symmetry remains unbroken. Our construction makes  use
only of fields contained in the  27-dimensional and singlet
representations of
$E_6$. All superpotential couplings are assumed to be generically of
order unity. The  mass spectrum entering the renormalization
group (RG) equations for $sin^2 \theta_w$ and $\alpha_s$ is completely
determined
by quadratic and cubic superpotential terms only.
This is  achieved  by an appropriate (minimal) choice of the  field  content.
Therefore, $M_c$ does not
enter the calculation  for $ sin^2\theta_w$ and $\alpha_s$ but is
only determined as
the scale where the unified $G$ gauge
structure constant $\alpha_G$ becomes equal to unity.

The left handed lepton, quark and antiquark superfields transform
under $G$ as follows:
\begin{displaymath}\begin{array}{lcccc}
\lambda & = & (1, \bar{3}, 3) & = & \left( \begin{array} {ccc} H^{(1)} &
H^{(2)} & L \\ E^c & \nu^c & N\end{array}\right),\strut \\

Q & =& (3,3,1) & = & \left ( \begin{array} {c}q\\ g\end{array}\right),
\strut \\

Q^c & = & (\bar{3}, 1, \bar{3}) & = &  \left( \begin{array} {c} u^c,
d^c, g^c\end{array}\right).\end{array}
\end{displaymath}

\noindent
Here $N$ and $\nu^c$ denote standard model singlet superfields while
$g(g^c)$ is an additional down-type quark (antiquark). We will be
working with eight fields of each type $(\lambda,Q,Q^c)$ and five
corresponding mirror fields $(\bar{\lambda},\bar{Q},\bar{Q}^c)$. Notice
that the field content is chosen to ensure identical running of the
three $SU(3)$ gauge couplings in the G-symmetric phase.

We impose invariance under three discrete  $Z_2$ symmetries
$P,C$ and $S$. Under $P$ all $\lambda, \bar{\lambda}$ fields remain
invariant while all $Q, \bar{Q},Q^c, \bar{Q}^c$ fields change sign.
This has the effect of forbidding all cubic superpotential couplings
involving three quark  or three antiquark fields.This fact combined with
the lack of gauge boson mediated proton decay leads to a practically
stable proton. Under C all fields remain invariant except $\lambda_5,
\lambda_8, \bar{\lambda}_3$ and $\bar{\lambda}_5$ which change sign.
Finally, under $S$ all fields remain invariant, except  $\lambda_5,
\lambda_6,\lambda_7, \bar{\lambda}_3, \bar{\lambda}_4$, $Q_1, Q_2$
and $Q^c_3$
which change sign.

The symmetry breaking of $G$ down to the standard group is obtained at
the scale $M_X\sim10^{16} GeV$ through appropriate superpotential
couplings of the fields acquiring a superlarge vacuum expectation
value. These are the fields
$\lambda_7,\bar{\lambda}_4,\lambda_8,\bar{\lambda}_5$. The superlarge
expectation values are:
$\><N_7>=$ $\>\><\bar{N}_4>^* =<\nu^c_8> = <\bar{\nu}^c_5>^*.$
We also introduce in the superpotential explicit mass terms of order
$M_X$ for all $Q,\bar{Q}$ and $Q^c,\bar{Q}^c $ pairs as well as all
$\lambda,\bar{\lambda}$ pairs involving fields which do not acquire
vacuum expectation values.

All vacuum expectation values leave the discrete symmetry P unbroken.
The discrete symmetry C combined with the $Z_2$ discrete subgroup
generated by the element $(diag (-1,-1), \> diag (-1,-1))$ of $SU(2)_L
\times SU(2)_R$ gives a $Z_2$ group $C'$ which remains unbroken by all the
vacuum expectation values and acts as " matter parity". Finally the
discrete symmetry S combined with the discrete subgroup generated by
the element $ diag(-1,+1,-1)$ of $SU(3)_R$ gives a $Z_2$ group $S'$
which remains unbroken by the superlarge expectation values. The role
of P in suppressing proton decay has been explained  earlier. The
"matter parity" $C'$ plays an essential role in suppressing some
lepton number violating couplings at the level of the MSSM. This is
important for the implementation of our mechanism for generating the
baryon asymmetry of the universe as well as in stabilizing the lowest
supersymmetric particle (LSP). Finally the combined effect of $C'$ and
$S'$ solves in a natural manner the gauge hierarchy problem.

We now turn to a brief discussion of the fermion mass spectrum below
the scale $M_X$. [The bosonic mass spectrum is directly obtainable
from the fermionic one, since supersymmetry is assumed to be exact
for energies much above $M_s \sim 1 TeV$.] A consequence of the fact
that the $Z_2$ symmetries $C'$ and $S'$ remain unbroken by the
superlarge vacuum expectation values is that all states can be
classified by their $C'$ and $S'$ charge. The mass matrices of the
various sectors are, therefore, broken up into submatrices  involving
only states characterized by common values  of $C'$ and $S'$. The
symmetry P leaves all mass terms invariant and therefore could not
have any consequences as far as the mass spectrum is concerned.

We will only describe in some detail the mass matrix of the $SU(2)_L$
-doublet leptons of positive matter parity $(C'=+1)$. These are the
fields with the correct quantum numbers to play the role of the
electroweak higgs doublets. Their mass matrix breaks up into two
submatrices characterized by the  $S'$-charge of the fields involved.
The submatrix of the $S' = +1$ sector couples  $H_k^{(2)} (k=1,...,4),
\bar{H}_4^{(1)} $ and $L_5$ with $H_6^{(1)}, H_7^{(1)},
\bar{H}_1^{(2)}, \bar{H}^{(2)}_2$ and $\bar{L}_3$. This is a
$6\times5$ matrix with 5 eigenvalues  $ \sim M_X$. One linear
combination, $h^{(2)}$, of the 6 fields, which  have the correct quantum
numbers to play the role of the  higgs doublet giving mass to ordinary
down-type quarks  and leptons, remains naturally exactly massless. The
corresponding $S'=-1$ submatrix couples $H^{(2)}_6,
H_7^{(2)},\bar{H}_1^{(1)},\bar{H}_2^{(1)}$ and $L_8$ with
$H^{(1)}_k (k=1,...,4),\bar{H}_4^{(2)}$ and $\bar{L}_5$. This is a
$5\times6$ matrix with 5 eigenvalues $\sim M_X$. Here again one
linear combination, $h^{(1)}$, of the 6 fields, which have the correct
quantum numbers to play the role of the higgs doublet giving mass to
ordinary up- type quarks, remains naturally exactly massless. We see
that the combined effect of the symmetries $C'$ and $S'$ solves in an
elegant manner the gauge hierarchy problem. The mechanism takes into
account all renormalizable as well as all the non-renormalizable
superpotential terms.

The states  which are not standard model singlets acquire masses $\sim
M_X$ except the three ordinary light generations all of which have
matter parity $C' = -1$. Two light lepton doublets as well as two
charged light antilepton singlets have negative $S'$-charge while the third
one has positive $S'$-charge. The same holds for the light quark
doublets as well as the light up- type antiquarks. In contrast, there
are two light down-type antiquarks with positive and only one with
negative $S'$ -charge. Taking into account the $C'$ and $S'$ -charges
of the light higgs doublets $h^{(1)},  h^{(2)}$, we conclude that one
light quark generation remains massless at tree-level.

As already emphasized, the fact that, below a unification scale $M_X$, we
recover the MSSM is sufficient to guarantee the successful RG
predictions for $sin^2 \theta_w$ and $\alpha_s$. Choosing the values
$M_s =1 $TeV and $M_X=1.3\times 10^{16}$ GeV, a one-loop calculation
gives the  perfectly
acceptable values $sin^2 \theta_w =0.230$ and $\alpha_s =0.116$ at the
electroweak scale.
Determining the scale $M_c$ by the
requirement that $\alpha_G(M_c) =1$, we obtain the value  $M_c=2.4
\times 10^{18}$ GeV which happens to be  equal to $M_P/\sqrt{8\pi}$.

The only standard model singlets that are of interest to us are the
negative matter parity ones with mass less than $M_X$. There are two
such states with positive and three with  negative $S'$ - charge. They
all have Majorana masses $\sim M_X^2/M_c$ from non-renormalizable
superpotential couplings  and play the role of the right-handed
neutrinos in the see-saw mechanism. The order of magnitude of their
mass allows for a solution of the solar neutrino puzzle through the
MSW effect$^{(4)}$ and at the same time does not exclude the possibility that
the
$\tau$-neutrino contributes significantly to the  missing mass of the
universe. A sizable right-handed neutrino mass allows the
implemention of
the mechanism of baryogenesis via leptogenesis.

By observing that the $S'$-charges of the electroweak higgs doublets
$h^{(1)}$ and $h^{(2)}$ are opposite we conclude that the  $S'$
symmetry cannot remain unbroken .
Actually, it must necessarily break by the  expectation value
of a standard model singlet in
order for the necessary mixing between $h^{(1)}$ and $h^{(2)}$ and a
mass term of the fermionic components  of $h^{(1)}$ and $h^{(2)}$ to
be generated. We assume that a gauge  singlet field T with positive
matter parity and negative $S'$ -charge acquires a vacuum expectation
value $\sim \frac {M_c}{M_X} M_s$ and a mass of the same order of
magnitude.
The necessary mixing $\sim M_s$ between
$h^{(1)}$ and $h^{(2)}$ will then be generated by the
non-renormalizable superpotential term $h^{(1)} h^{(2)} N_7 T/M_c$.
This term upon substitution of $N_7$ with its vacuum expectation
value represents the dominant coupling between  T and the MSSM
states.
Notice that the spontaneous breaking of the  discrete symmetry $S'$
does not lead to the well-known domain wall cosmological problem
because of the $SU(3)_c$ -anomaly of the discrete symmetry $S'^{(5)}$.

One could obtain a more natural mechanism for generating  a non-zero
higgsino mass involving only scales $ \sim M_X$ at the  expense of
introducing a $Z_5$ symmetry acting non-trivially only on gauge
singlets. We introduce two such gauge singlets T and  $\overline{T}$
both with positive matter parity and negative  $S'$ -charge on
which the new symmetry generator acts as $\alpha = exp(2 \pi i/5)$ and
$\alpha^4$ respectively. These singlets can acquire naturally
expectation values  $\sim M_X$ thereby breaking the $S'$
symmetry at a superheavy scale. Remarkably enough, the higgsino mass
generated by the non-renormalizable terms $h^{(1)} h^{(2)} N_7
T^5/M^5_c$ and  $h^{(1)} h^{(2)} N_7 \overline{T}^5/M^5_c$ is still
acceptably small whereas the singlets T and $\overline{T}$ are obviously
sufficiently decoupled. Discrete symmetries  broken spontaneously at
superlarge scales $\sim M_X$ are not expected  to give rise to domain
wall formation in the inflationary universe scenario.

We conclude by summarizing our results. We presented a scheme in which
a $G\equiv SU(3)^3$ supersymmetric gauge theory with one gauge
coupling  $\alpha_G=1$ and symmetric spectrum among the three $SU(3)$
gauge groups arises as an effective  4-dimensional theory at a scale
$M_c=M_P/\sqrt{8\pi}$. Spontaneous breaking of this gauge group at a
scale $M_X=1.3\times 10^{16}$GeV gives naturally rise exactly to the
MSSM, a result which guarantees as a prediction the experimentally
favored values of $sin^2\theta_w$ and $\alpha_s$. We relied on the
combined effect of three $Z_2$ discrete symmetries, one to stabilize the
proton, the second to act like a "matte parity" and the third to solve
the gauge hierarchy problem. We easily obtain two
neutrinos light enough to allow for a solution of the solar neutrino problem
through
the MSW effect and  a third neutrino which possibly gives a
considerable contribution to the
"hot" component of the dark matter of the universe. The LSP is stable
and could be the dominant "cold" component of dark matter. Lepton number
asymmetry is generated through right-handed neutrino decays and is
later partially transformed into baryon asymmetry by the
non-perturbative baryon-
and -lepton-number-violating effects of the standard electroweak
theory.
Finally,
domain wall formation is avoided. We believe that models of this
type should be seriously considered as viable alternatives to the
usual minimal supersymmetric GUTs.
\newpage
\section*{References}

\begin{enumerate}
\item  J. Ellis, S.Kelley and D.V.Nanopoulos, Phys.Lett.\underline
{B249} (1990)
441;
U.Amaldi, W. de Boer and H. Furstenan, Phys. Lett. \underline{B260}
(1991) 447;
P.Langacker and M.X.Luo, Phys. Rev. \underline {D44} (1991) 817.

\item S. Dimopoulos and H. Georgi, Nucl. Phys. \underline {B193}
(1981)150;
N. Sakai, Z. Phys. \underline {C11} (1981) 153.

\item G. Lazarides, C. Panagiotakopoulos and Q. Shafi, Phys. Lett.
\underline {B315} (1993) 325; E.\underline{B317} (1993)661.

\item L. Wolfenstein, Phys. Rev. \underline {D17} (1978) 2369;
Phys. Rev. \underline {D20} (1979) 2634;
S. P. Mikheyev and A. Yu. Smirnov, Usp. Fiz. Nauk. \underline {153}
(1987) 3.

\item J. Preskill, S.P. Trivedi, F. Wilczek and M.B. Wise, Caltec
preprint CALT-68-1718 (1991).

\end{enumerate}
\end{document}